\documentclass[pra,twocolumn,floatfix,aps,superscriptaddress,showpacs,amsfonts]{revtex4}
\usepackage[T1]{fontenc}
\usepackage{graphicx}
\usepackage{array}
\usepackage{amsthm}
\usepackage{amsmath}
\usepackage{amssymb}
\usepackage{xypic}

\begin{document}
\title{The one-way unlocalizable quantum discord}

\author{Zhengjun Xi}
\email{snnuxzj@gmail.com}
\affiliation{College of Computer Science, Shaanxi Normal University, Xi'an, 710062, China}

\author{Heng Fan}
\affiliation{Institute of Physics, Chinese Academy of Sciences, Beijing 100190, China}

\author{Yongming Li}
\affiliation{College of Computer Science, Shaanxi Normal University, Xi'an, 710062, China}

\date{\today}

\begin{abstract}
In this paper, we present the concept of the one-way unlocalizable quantum discord and investigate its properties.
We provide a polygamy inequality for it in tripartite pure quantum system of arbitrary dimension.
Several tradeoff relations between the one-way unlocalizable quantum discord and other correlations are given.
If the von Neumann measurement is on a part of the system, we give two expressions of the one-way unlocalizable quantum discord
in terms of partial distillable entanglement and quantum disturbance. Finally, we also provide a lower bound for bipartite shareability of quantum correlation beyond entanglement in a tripartite system.
\end{abstract}
\eid{identifier}
\maketitle
\section{Introduction}
Entanglement plays a key role in quantum computation and quantum information, which
is a kind of quantum correlation not available in classical world \cite{HorodeckiRMP}.
On the other hand, there exists some other quantum correlations different from
entanglement, for example, quantum discord is also a quantum correlation not
present in classical system \cite{Ollivier02,HendersonJPA}.
It is supposed that quantum discord might be responsible for the advantage of a
quantum algorithm, the deterministic quantum computation
with one qubit \cite{Knill}, which surpasses any corresponding classical algorithms while
with vanishing entanglement but non-zero quantum discord. It is also noticed that
though different from entanglement, quantum discord is closely related with
entanglement and may have interesting operational interpretations in quantum
information processing
\cite{Cavalcanti-fan,Madhok-fan,Streltsov11,Piani11,SKBnew,MazzolaNSR11,Campbell11,Ciccarello5551}, and also see the reviews~\cite{discordreview,Celeri11}.

We know that entanglement is an
invaluable resource for quantum information processing, and thus it
cannot be shared freely among multiparities. This property is described
by monogamy inequality which is a
one of the most fundamental properties of entanglement~\cite{HorodeckiRMP,CKWPRA}.
That is, if two qubits $A$ and $B$ are maximally entangled
they cannot be correlated at all with third qubit $C$. In
general, there is tradeoff between the amount of entanglement between qubits $A$ and $B$ and the same qubit $A$
and qubit $C$. This property is purely quantum: in classical world if $A$ and $B$ bits are perfectly correlated, then
there is no constraints on correlations between bits $A$ and
$C$~\cite{HorodeckiRMP}. For three qubits the tradeoff is described by Coffman-Kundu-Wootters (CKW) monogamy inequality in terms of the concurrence~\cite{CKWPRA}.
The CKW inequality can be extended to $n$-qubits, it has
been proven true only recently~\cite{OsbornePRL}.
More discussions on monogamy of different quantum correlations measures can
be found in~\cite{Seevinck10,Kay09,Paw09,Kurzynski11,Ou07,Ou08}.

Since quantum
discord quantifies the quantum correlation in a bipartite
state~\cite{Ollivier02,HendersonJPA} and might also be a resource in
quantum information processing, it is interesting to study whether it also respects
monogamy relation. Recently, monogamy relation for quantum discord has already been
studied in \cite{Prabhu,Giorgi11,Ren11,Rajagopal,StreltsovarXiv11}.
It is found that monogamy of quantum discord is due to the relationships between the entanglement of formation and the one-way classical correlation, see the inequality~(\ref{eq:EoF_QD}). In general, this inequality (\ref{eq:EoF_QD}) does not hold~\cite{LuoPRAa}. Thus, the monogamy of quantum discord is not universal for any tripartite pure state. As pointed out in~\cite{Prabhu},
polygamy relation for quantum discord can hold for some states.

For the case of entanglement, it is showed that distributed entanglement is by nature polygamy by establishing the concept
of the one-way unlocalizable entanglement \cite{BuscemiPRA}, which is defined by entanglement of assistance~\cite{Divincenzo99,Cohen98}.
Since quantum discord is an important supplementary to entanglement, it
motivates us to propose a similar concept for quantum discord.
Thus we can systematically investigate the polygamy of distributed quantum correlations in terms of quantum discord in tripartite quantum system.
Based on the duality between the one-way unlocalizable entanglement and the one-way classical correlation~\cite{HendersonJPA},
in this paper, we present the concept of the one-way unlocalizable quantum discord and provide an operational interpretation.
We give the lower and upper bound for the one-way unlocalizable quantum discord. We also derive several tradeoff relations between the one-way unlocalizable quantum discord and other correlations, and provide a polygamy relation of distributed quantum correlations in terms of the one-way unlocalizable quantum discord.
In case the von Neumann measurement is on a part of the system,
we give two expressions of the one-way unlocalizable quantum discord in terms of partial distillable entanglement~\cite{Streltsov11} and quantum disturbance~\cite{Buscemi08}. Finally, we also provide a lower bound for bipartite shareability of quantum correlations (beyond entanglement) in a tripartite system.

The paper is organized as follows. In Sec.\ref{sec:1way UQD}, we provide the definition of the one-way unlocalizable quantum discord and investigate its basic properties, and give a polygamy relation of distributed quantum correlations. In Sec.\ref{sec:linking QC and Entanglement}, we study the relations between the one-way unlocalizable quantum discord and entanglement.  In Sec.\ref{sec:monogamy of QD}, we discuss monogamy relation for quantum discord
 and we summarize our results in Sec. \ref{sec:conclusion}.

\section{One-way unlocalizable quantum discord}\label{sec:1way UQD}
\subsection{Definition}

We now consider a tripartite pure state $|\psi\rangle^{ABC}$ shared between three parties refereed to as Alice, Bob and Charlie. The entanglement supplier, Bob preforms a measurement on him share of the tripartite state which yields a know bipartite entangled state for Alice and Charlie. Tracing over Bob's system yields the bipartite mixed state
\begin{equation}
\rho^{AC}=\mathrm{Tr}_B(|\psi\rangle^{ABC}\langle\psi|)\label{eq:reduced state_AC}
\end{equation}
shared by Alice and Charlie~\cite{HughstonPLA,GourPRA}.

Bob's aim is to maximize entanglement for Alice and Charlie, and the maximum average entanglement he can create is the entanglement of assistance, which was originally defined in terms of entropy of entanglement. As a dual quantity to entanglement of formation~\cite{Wootters98,Wootters01}, entanglement of assistance~\cite{Divincenzo99,Cohen98} is defined by the maximum average entanglement  of $\rho^{AC}$
\begin{equation}
E_a(\rho^{AC})=\max_{\{p_x,|\phi_x\rangle^{AC}\}}\sum_xp_xS(\rho_x^A),\label{eq:EOA}
\end{equation}
where the maximum is taken as over all possible pure state decompositions of $\rho^{AC}$, satisfying $\rho^{AC}=\sum_xp_x|\phi_x\rangle^{AC}\langle\phi_x|$ and $\rho_x^A=\mathrm{Tr}_C(|\phi_x\rangle^{AC}\langle\phi_x|)$. Here, $S(\rho):=-\mathrm{Tr}\rho\log\rho$ is von Neumann entropy~\cite{Nielsen}.

Since $E_a(\rho^{AC})$ measures the maximum average entanglement that can be localized on the subsystem $AC$ with the assistance of $B$, and $E_a(\rho^{AC})\leq S(\rho^{A})$, then the difference between two is the so-called the one-way unlocalizable entanglement of a bipartite state $\rho^{AB}$, namely,
\begin{equation}
E_u^{\leftarrow}(\rho^{AB})=S(\rho^A)-E_a(\rho^{AC}),\label{eq:BGKa10}
\end{equation}
where $\rho^{AB}=\mathrm{Tr}_C(|\psi\rangle^{ABC}\langle\psi|)$~\cite{BuscemiPRA}.

For tripartite pure state $|\psi\rangle^{ABC}$, all possible pure state decomposition of $\rho^{AC}$ can be realized by rank-1 measurements of subsystem $B$, conversely, any rank-1 measurement can be induced from a pure state decomposition of $\rho^{AC}$~\cite{HughstonPLA,GourPRA,Preskilllecture,BuscemiPRA}. Therefore, the right-hand side of Eq.(\ref{eq:EOA}) is the maximum average entropy overall possible rank-1 measurements $\{M^B_k\}$ applied on the subsystem $B$. Then, Eq.(\ref{eq:BGKa10}) can be equivalently characterized by
\begin{equation}
E_u^{\leftarrow}(\rho^{AB})=\min_{\{M^B_k\}}\Big[S(\rho^A)-\sum_kp_kS(\rho_k^A)\Big],\label{eq:one-way UE1}
\end{equation}
where the minimum is taken over all possible rank-1 measurements $\{M^B_k\}$ applied on subsystem $B$, $p_k=\mathrm{tr}[(I^A\otimes M^B_x)\rho^{AB}]$ is the probability of the outcome $k$, and $\rho_k^{A}=\mathrm{Tr}_B[(I^A\otimes M^B_k)\rho^{AB}]/p_k$ is the state of system $A$ when the outcome is $k$~\cite{BuscemiPRA}.

At the same time, the one-way classical correlation~\cite{HendersonJPA} is defined as
\begin{equation}
\mathcal{J}^{\leftarrow}(\rho^{AB})=\max_{\{M^B_k\}}\Big[S(\rho^A)-\sum_kp_kS(\rho_k^A)\Big].\label{eq:one-way CC}
\end{equation}
The relation between the one-way unlocalizable entanglement and the one-way classical correlation is like the relation between the entanglement of assistance and the entanglement of formation.

The quantum mutual information is a measure of the total amount of
correlations in the bipartite quantum state, and it was divided into quantum correlation and classical correlation. Then, the one-way unlocalizable quantum discord of the state $\rho^{AB}$ under the measurement elements $\{M^B_k\}$ is defined as the difference between the mutual information and the one-way unlocalizable entanglement, namely,
\begin{equation}
\delta_u^{\leftarrow}(\rho^{AB}):=\mathcal{I}(\rho^{AB})-E_u^{\leftarrow}(\rho^{AB}),\label{eq:QD_uAB}
\end{equation}
where $\mathcal{I}(\rho^{AB}):= S(\rho^{A})+S(\rho^{B})-S(\rho^{AB})$ is mutual information~\cite{Nielsen}.
This quantity is inspired by the definition of quantum discord~\cite{Ollivier02}, in some sense dual to quantum discord, and it is defined as
\begin{equation}
\delta^{\leftarrow}(\rho^{AB}):=\mathcal{I}(\rho^{AB})-\mathcal{J}^{\leftarrow}(\rho^{AB}).\label{eq:QDA_B}
\end{equation}

\subsection{Tradeoff relation}

\begin{figure}
  \includegraphics[scale=0.3]{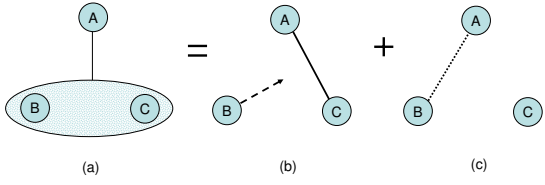}
  \caption{(Color online) The entanglement of $|\psi\rangle^{A(BC)}$ (a) consists of two parts: (a) the localizable entanglement $E_a(\rho^{AC})$ and (b) the unlocalizable entanglement $E_u^{\leftarrow}(\rho^{AB})$.This figure is reproduced from~\cite{BuscemiPRA}.\label{fig:1}}
\end{figure}

To understand the one-way unlocalizable quantum discord, we firstly give an operational interpretation. All bipartite and single-party states are obtained by taking the appropriate partial traces of tripartite pure state. Then, without loss of generally, we suppose that $|\psi\rangle^{ABC}$ is the purification of $\rho^{AB}$.
In this case, Eq.(\ref{eq:BGKa10}) can be rewritten as
\begin{subequations}
\begin{equation}
S(\rho^A)=E_u^{\leftarrow}(\rho^{AB})+E_a(\rho^{AC}),\label{eq:BGKa1}
\end{equation}
where $S(\rho^A)$ quantifies the entanglement of the pure state $|\psi\rangle^{A(BC)}$ with respect to the bipartite cut $A-BC$. This equality is called the Buscemi-Gour-Kim equality. Fig.\ref{fig:1} graphically illustrates this separation, we follow this figure from~\cite{BuscemiPRA}.
In fact, the Buscemi-Gour-Kim equality (\ref{eq:BGKa1}) is universal for any tripartite pure state, namely,
\begin{equation}
S(\rho^X)=E_u^{\leftarrow}(\rho^{XY})+E_a(\rho^{XZ}),\label{eq:BGKc2}
\end{equation}
where $X, Y, Z\in \{A,B, C\}$.
\end{subequations}

For the tripartite pure state $|\psi\rangle^{ABC}$, using the Buscemi-Gour-Kim equality (\ref{eq:BGKa1}), we have
\begin{equation}
\delta_u^{\leftarrow}(\rho^{AB})+S(A|B)=E_a(\rho^{AC}).\label{eq:QD_u_EOA1}
\end{equation}
where $S(A|B):=S(\rho^{AB})-S(\rho^B)$ is the von Neumann conditional entropy~\cite{Nielsen}. Since $S(\rho^C)=S(\rho^{AB})$, using Eq.(\ref{eq:BGKc2}), we obtain the tradeoff relation between the one-way unlocalizable quantum discord for $A$ and $B$ and the one-way unlocalizable entanglement for $C$ and $B$, namely,
\begin{equation}
S(\rho^{B})=\delta_u^{\leftarrow}(\rho^{AB})+E_u^{\leftarrow}(\rho^{CB}).\label{eq:UQD_UE}
\end{equation}

\begin{figure}
  \includegraphics[scale=0.3]{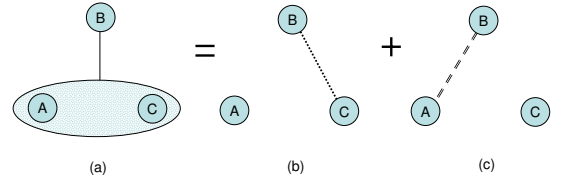}
  \caption{(Color online) The entanglement of $|\psi\rangle^{B(AC)}$ (a) consists of two parts: (b) the unlocalizable entanglement $E_u^{\leftarrow}(\rho^{BC})$ and (c) the one-way unlocalizable quantum discord $\delta_u^{\leftarrow}(\rho^{AB})$.
\label{fig:2}}
\end{figure}

Here, $S(\rho^B)$ quantifies the entanglement of the pure state $|\psi\rangle^{B(AC)}$ with respect to the bipartite cut $B-AC$. This equation says that it is equal to the sum between the one-way unlocalizable quantum discord for $A$ and $B$ with measurements on $B$ and the one-way unlocalizable entanglement for $C$ and $B$ with measurements on $B$.
We graphically illustrate this separation in Fig.\ref{fig:2}.

We now list some basic properties of the one-way unlocalizable quantum discord.
\begin{enumerate}
\item For any bipartite state $\rho^{AB}$,
\begin{equation}
\frac{\mathcal{I}(\rho^{AB})}{2}\leq\delta_u^{\leftarrow}(\rho^{AB})\leq S(\rho^B).
\end{equation}
In particular, for any pure state $|\psi\rangle^{AB}$, we have
\begin{equation}
\delta_u^{\leftarrow}(\rho^{AB})=E_u^{\leftarrow}(\rho^{AB})=S(\rho^B).\label{eq:UQD pure}
\end{equation}
\item For all bipartite states $\rho^{AB}$ and $\rho^{A'B'}$,
\begin{equation}
\delta_u^{\leftarrow}(\rho^{AB})+\delta_u^{\leftarrow}(\rho^{A'B'})\leq\delta_u^{\leftarrow}(\rho^{AB}\otimes\rho^{A'B'}).
\end{equation}
\item For a three-qubit pure state $|\psi\rangle^{ABC}$,
\begin{equation}
S(\rho^{A})\leq\delta_u^{\leftarrow}(\rho^{AB})+\delta^{\leftarrow}(\rho^{AC}).
\end{equation}
\end{enumerate}

\begin{proof}
1. For any bipartite state $\rho^{AB}$, form the result~\cite{BuscemiPRA}, one obtain
\begin{equation}
E_u^{\leftarrow}(\rho^{AB})\leq \frac{\mathcal{I}(\rho^{AB})}{2}.\label{eq:UE_upper bound}
\end{equation}
Substituting this equation into Eq.(\ref{eq:QD_uAB}), we obtain a lower bound
\begin{equation}
\frac{\mathcal{I}(\rho^{AB})}{2}\leq\delta_u^{\leftarrow}(\rho^{AB}).
\end{equation}
Since $E_u^{\leftarrow}(\rho^{AB})\geq 0$, from Eq.(\ref{eq:UQD_UE}), we give the upper bound
\begin{equation}
\delta_u^{\leftarrow}(\rho^{AB})\leq S(\rho^B).
\end{equation}

For pure state case, since every post-measurement state $\rho^{A}_k$ is also pure state, then Eq.(\ref{eq:UQD pure}) can be directly verified.

2. For all bipartite states $\rho^{AB}$ and $\rho^{A'B'}$, we have
\begin{align}
&\delta_u^{\leftarrow}(\rho^{AB}\otimes\rho^{A'B'})\nonumber\\
=&\mathcal{I}(\rho^{AB}\otimes\rho^{A'B'})-E_u^{\leftarrow}(\rho^{AB}\otimes\rho^{A'B'})\nonumber\\
\geq& \mathcal{I}(\rho^{AB})+\mathcal{I}(\rho^{A'B'})-E_u^{\leftarrow}(\rho^{AB})-E_u^{\leftarrow}(\rho^{A'B'})\nonumber\\
=&\delta_u^{\leftarrow}(\rho^{AB})+\delta_u^{\leftarrow}(\rho^{A'B'}),
\end{align}
where the inequality is due to the additivity of quantum mutual information~\cite{Nielsen} and the subadditivity of the one-way unlocalizable entanglement~\cite{BuscemiPRA}.

3. Following the results~\cite{BuscemiPRA}, for a three-qubit pure state $|\psi\rangle^{ABC}$,
\begin{equation}
S(\rho^{A})-E_a(\rho^{AC})\leq E_f(\rho^{AB}).
\end{equation}
It is actually equivalent to
\begin{equation}
E_u^{\leftarrow}(\rho^{AB})\leq E_f(\rho^{AB}).
\end{equation}
Using the Koashi-Winter equality~(\ref{eq:KW}), we have
\begin{equation}
E_u^{\leftarrow}(\rho^{AB})\leq S(\rho^A)-\mathcal{J}^{\leftarrow}(\rho^{AC}).
\end{equation}
After some manipulation we obtain
\begin{equation}
S(\rho^A)\leq \delta_u^{\leftarrow}(\rho^{AB})+\delta^{\leftarrow}(\rho^{AC}).
\end{equation}
\end{proof}

\subsection{Polygamy inequality}
Since the Buscemi-Gour-Kim equality (\ref{eq:BGKa1}) is universal for any tripartite pure state, one checks that
\begin{equation}
\delta_u^{\leftarrow}(\rho^{AC})+S(A|C)=E_a(\rho^{AB}).\label{eq:QD_u_EOA2}
\end{equation}
Combined Eq.(\ref{eq:QD_u_EOA1}), we obtain the conservation relation between the one-way unlocalizable quantum discord and the entanglement of assistance as follows
\begin{equation}
E_a(\rho^{AC})+E_a(\rho^{AB})=\delta_u^{\leftarrow}(\rho^{AB})+\delta_u^{\leftarrow}(\rho^{AC}).\label{eq:assistance_conservation}
\end{equation}

From the results in~\cite{BuscemiPRA}, there exists a polygamous relation of entanglement of assistance, namely,
\begin{equation}
E_a(\rho^{A(BC)})\leq E_a(\rho^{AB})+E_a(\rho^{AC)}).\label{eq:poly_EOA}
\end{equation}
The one-way unlocalizable quantum discord is equal to the entanglement of assistance for the pure state $|\psi\rangle^{ABC}$ with respect to the bipartite cut $A-BC$, we then have
\begin{equation}
\delta_u^{\leftarrow}(\rho^{A(BC)})=E_a(\rho^{A(BC)}).\label{eq:UQD equal EoA}
\end{equation}
Substituting Eqs.(\ref{eq:poly_EOA}) and (\ref{eq:UQD equal EoA}) into Eq.(\ref{eq:assistance_conservation}), we get the polygamy of the one-way unlocalizable quantum discord
\begin{equation}
\delta_u^{\leftarrow}(\rho^{A(BC)})\leq\delta_u^{\leftarrow}(\rho^{AB})+\delta_u^{\leftarrow}(\rho^{AC}).\label{eq:poly_UQD}
\end{equation}
This inequality show that there exists a polygamy relation of quantum correlations in terms of the one-way unlocalizable quantum discord for a tripartite pure state of arbitrary dimension.

\section{Linking quantum correlation to entanglement}\label{sec:linking QC and Entanglement}

\begin{figure}
  \includegraphics[scale=0.4]{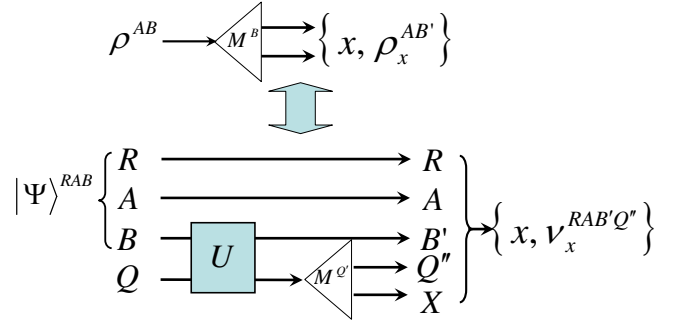}
  \caption{(Color online) The action of a quantum measurement $M^B$ on the
input state $\rho^{AB}$, can always be extended as a tripartite indirect measurement (bottom), where the apparatus $Q$, after
having properly interacted with the input system through $U^{BQ}$,
undergoes the measurement $M^{Q'}$ . The conditional output pure states $\{\upsilon^{RAB'Q''}_x\}_x$ are such that $\mathrm{Tr}_{RQ''}(\upsilon^{RAB'Q''}_x)=\rho^{AB'}_x$. This figure is reproduced from~\cite{Buscemi08}.
\label{fig:3}}
\end{figure}

First we introduce the measures of information gain
and disturbance of a quantum measurement constructed
in~\cite{Buscemi08}. Form the result~\cite{Ozawa84}, it states that whatever quantum
measurement $\{M^B_x\}$ can be modeled as an indirect measurement,
in which the input system first interacts with an
apparatus $Q$, initialized in a fixed pure state
$|0\rangle^Q$, through a suitable unitary interaction $I^A\otimes U^{BQ}: \mathcal{H}_{ABQ}\rightarrow \mathcal{H}_{AB'Q'}$, and
\begin{equation}
\Big(I^A\otimes U^{BQ}\Big)\Big(\rho^{AB}\otimes |0\rangle^Q\langle 0|\Big)\Big(I^A\otimes U^{\dagger BQ}\Big)=\rho^{AB'Q'}.\nonumber
\end{equation}
Subsequently, a particular measurement $M^{Q'}$, depending also on $U^{BQ}$, is performed on the apparatus.
In addition, by introducing a reference system $R$ purifying the input state as $|\Psi\rangle^{RAB}$, $\mathrm{Tr}_R(|\Psi\rangle^{RAB}\langle\Psi|)=\rho^{AB}$. we are in the situation schematically represented as in Fig.\ref{fig:3}.
After the unitary interaction $U^{BQ}$, the global state is
\begin{equation}
|\Upsilon\rangle^{RAB'Q'}=\Big(I^{RA}\otimes U^{BQ}\Big)|\Psi\rangle^{RAB}\otimes|0\rangle^Q,\nonumber
\end{equation}
and the measurement on the apparatus can be chosen such that
\begin{equation}
(I^{RAB'Q'}\otimes M^{Q''})|\Upsilon\rangle^{RAB'Q'}=\sum_xp_x\upsilon_x^{RAB'Q''}\otimes|x\rangle^X\langle x|,\nonumber
\end{equation}
where $\upsilon_x^{RAB'Q''}$ are pure states such that
\begin{equation}
\mathrm{Tr}_{Q''}(\upsilon_x^{RAB'Q''})=(I^{RA}\otimes M^{B}_x)|\Psi\rangle^{RAB}/p_{x}=\rho^{RAB'}\nonumber
\end{equation}
and
\begin{equation}
\mathrm{Tr}_R\rho^{RAB'}=\rho^{AB'}_x=(I^A\otimes E^B_x)\rho^{AB}(I^A\otimes E^{\dagger B}_x),\nonumber
\end{equation}
where $|x\rangle^X$ are the classical register states, $M^B_x:=E^{\dagger B}_xE^B_x$.

We note that the one-way unlocalizable quantum discord is defined by considering rank-1 POVMs. The rank-1 POVMs can be replaced by a minimization over orthogonal projectors of rank-1 on an extended Hilbert space~\cite{Peresbook}. Here, the generalized measurement are actually restricted to complete von Neumann measurements.

 Based on the result in~\cite{Streltsov11}, we establishes a connection between the partial entanglement and the one-way unlocalizable quantum discord. The one-way unlocalizable quantum discord of a bipartite state $\rho^{AB}$ is equal to the maximal partial distillable entanglement in a von Neumann measurement on $B$,
 \begin{equation}
 \delta_u^{\leftarrow}(\rho^{AB})=\max_{U}\Big[E^{AB'|Q'}_D(\tilde{\rho}^{AB'M})-E^{B'|Q'}_D(\tilde{\rho}^{B'Q'})\Big],
 \end{equation}
 where the maximum is taken over unitaries $U$ which realize a von Neumann measurement on $B$.
Here, $E_D$ is distillable entanglement~\cite{Plenio07,Bennett96}. For arbitrary measurement of entanglement $E$, one denotes the partial entanglement
 \begin{equation}
 P_E:=E^{AB'|Q'}(\tilde{\rho}^{AB'Q'})-E^{B'|Q'}(\tilde{\rho}^{B'Q'}).\nonumber
 \end{equation}
 It quantifies the part of entanglement which is lost when the subsystem $B$ is ignored~\cite{Streltsov11}.

Under the measurement $M^B$ on $B$, the quantum disturbance introduced $\delta(\rho^B,M^B)$ and $\delta(\rho^{AB},M^B)$ are given, respectively,
\begin{subequations}
\begin{equation}
\delta(\rho^B,M^B):=S(\rho^B)-\mathcal{I}_{coh}^{RA\rightarrow B'X}(\rho^{RAB'X}),\label{eq:quantum disturbance_1}
\end{equation}
\begin{equation}
\delta(\rho^{AB},M^B):=S(\rho^{AB})-\mathcal{I}_{coh}^{R\rightarrow AB'X}(\rho^{RAB'X}),\label{eq:quantum disturbance_2}
\end{equation}
\end{subequations}
where $\mathcal{I}_{coh}^{Y\rightarrow YZ}:=S(\rho^{Z})-S(\rho^{YZ})$ is the coherent
information and $R$ is a purifying system for which $|\Psi\rangle^{RAB}$
is the purified state of density operator $\rho^{AB}$. Here, $\rho^{RAX}$ and
$\rho^{RX}$ are the states reduced from the post measurement
state $\rho^{RAB'X}$.

After some manipulation one obtain the following form
\begin{align}
&\delta(\rho^B,M^B)-\delta(\rho^{AB},M^B)\nonumber\\
=&S(\rho^{B})-S(\rho^{AB})+\sum_xp_xS(\rho^A_x).
\end{align}
Therefore, we obtain the second expressions for the one-way unlocalizable quantum discord based on quantum disturbance, namely,
\begin{equation}
 \delta_u^{\leftarrow}(\rho^{AB})=\max_{M^B}\left[\delta(\rho^B,M^B)-\delta(\rho^{AB},M^B)\right].
\end{equation}

At the same time, the
information gain $\iota(\rho^{B},M^B)$ and $\iota(\rho^{AB},M^B)$ are given, respectively,
\begin{subequations}
\begin{equation}
\iota(\rho^{B},M^B):=\mathcal{I}^{RA:X}(\rho^{RAX}),\label{eq:information gain_1}
\end{equation}
\begin{equation}
\iota(\rho^{AB},M^B):=\mathcal{I}^{R:X}(\rho^{RX}).\label{eq:information gain_2}
\end{equation}
\end{subequations}
After some manipulation one obtain the following form
\begin{align}
&\iota(\rho^B,M^B)-\iota(\rho^{AB},M^B)\nonumber\\
=&S(\rho^{B})-S(\rho^{AB})+\sum_xp_x\Big[S(\rho^R_x)-S(\rho^{RA}_x)\Big]\nonumber\\
=&S(\rho^{B})-S(\rho^{AB})+\sum_xp_x\Big[S(\rho^{AB'Q''}_x)-S(\rho^{B'Q''}_x)\Big]\nonumber\\
\leq &S(\rho^{B})-S(\rho^{AB})+\sum_xp_xS(\rho^A_x).
\end{align}
The second equality is due to the $\upsilon_x^{RAB'Q''}$ is pure state for every $x$. The inequality come form the sub-additivity of the von Neumann entropy. Therefore, we give another lower bound for the one-way unlocalizable quantum discord based on information gain, namely,
\begin{equation}
\delta_u^{\leftarrow}(\rho^{AB})\geq \max_{M^B}\left[\iota(\rho^B,M^B)-\iota(\rho^{AB},M^B)\right]
\end{equation}
with equality if and only if $S(\rho^{RA}_x)=S(\rho^R_x)-S(\rho^A_x)$ for every $x$.

\section{Monogamy of quantum discord}\label{sec:monogamy of QD}

 For the tripartite pure state $|\psi\rangle^{ABC}$ with $\rho^A=\mathrm{Tr}_{BC}(|\psi\rangle^{ABC}\langle\psi|)$, $\rho^{AB}=\mathrm{Tr}_{C}(|\psi\rangle^{ABC}\langle\psi|)$ and $\rho^{AC}=\mathrm{Tr}_{B}(|\psi\rangle^{ABC}\langle\psi|)$, the Koashi-Winter equality is given as~\cite{KoashiPRA}
\begin{equation}
E_f(\rho^{AB})+\mathcal{J}^{\leftarrow}(\rho^{AC})=S(\rho^A).\label{eq:KW}
\end{equation}
Here $\mathcal{J}^{\leftarrow}(\rho^{AC})$ is the one-way classical correlation between $A$ and $C$, and
the entanglement of formation $E_f$ is defined as
\begin{equation}
E_f(\rho^{AB})=\min_{\{p_i,|\psi_i\rangle^{AB}\}}\sum_ip_iS(\rho_i^A),\label{eq:EOF}
\end{equation}
where the minimum is taken over all pure ensembles $\{p_i,|\psi_i\rangle\}$
satisfying $\rho^{AB}=\sum_ip_i|\psi_i\rangle^{AB}\langle\psi_i|$, and $\rho_i^A=\mathrm{Tr}_B(|\psi_i\rangle^{AB}\langle\psi_i|)$~\cite{Wootters98,Wootters01}.

For the tripartite pure state $|\psi\rangle^{ABC}$, the monogamic distribution of entanglement of formation and quantum discord can be given as~\cite{Fanchini11a}
\begin{equation}
\delta^{\leftarrow}(\rho^{AB})+\delta^{\leftarrow}(\rho^{AC})=E_f(\rho^{AB})+E_f(\rho^{AC}).\label{eq:conservation relation}
\end{equation}
Here, $\delta^{\leftarrow}(\rho^{XY})$ is the quantum discord of $\rho^{XY}$.

If the entanglement of formation would be smaller than the one-way classical correlation, namely,
\begin{equation}
E_f(\rho^{XY})\leq \mathcal{J}^{\leftarrow}(\rho^{XY}).\label{eq:EoF_QD}
\end{equation}
Substituting this inequality into Eq.(\ref{eq:KW}), there exists a monogamy relation of entanglement of entanglement for the tripartite pure states. Along with this way, combining Eq.(\ref{eq:conservation relation}), we give the monogamy of quantum discord as following
\begin{equation}
\delta^{\leftarrow}(\rho^{AB})+\delta^{\leftarrow}(\rho^{AC})\leq \delta^{\leftarrow}(\rho^{A(BC)}),\label{eq:monogamy of QD}
\end{equation}
where $\delta^{\leftarrow}(\rho^{A(BC)})=E_f(\rho^{A(BC)})=S(\rho^A)$, and $S(\rho^A)$ quantifies the entanglement of the pure state $|\psi\rangle^{ABC}$ with respect to the bipartite cut $A-BC$.

But the inequality (\ref{eq:EoF_QD}) does not hold for some states~\cite{LuoPRAa,HendersonJPA}. Thus, the monogamy of quantum discord is not universal for tripartite any pure state (see~\cite{Giorgi11,Rajagopal,Ren11,Streltsov11}).

Combined Eq.(\ref{eq:BGKa1})with Eq.(\ref{eq:KW}), we obtain the conservation relation
\begin{equation}
E_f(\rho^{AB})+\mathcal{J}^{\leftarrow}(\rho^{AC})=E_a(\rho^{AC})+E_u^{\leftarrow}(\rho^{AB}).
\end{equation}

It is easy to verify that the the one-way unlocalizable quantum discord is equal to quantum discord for the pure state $|\psi\rangle^{ABC}$ with
respect to the bipartite cut $A-BC$. By the definitions of entanglement of formation and entanglement of assistance, we have
\begin{equation}
E_f(\rho^{XY})\leq E_a^{\leftarrow}(\rho^{XY}).
\end{equation}
Using Eq.(\ref{eq:conservation relation}), we obtain an interesting inequality as follows
\begin{align}
\delta^{\leftarrow}(\rho^{AB})+\delta^{\leftarrow}(\rho^{AC})&=E_f(\rho^{AB})+E_f(\rho^{AC})\nonumber\\
&\leq E_a^{\leftarrow}(\rho^{AB})+E_a^{\leftarrow}(\rho^{AC})\nonumber\\
&=\delta_u^{\leftarrow}(\rho^{AB})+\delta_u^{\leftarrow}(\rho^{AC}).\label{eq:UQD and QD}
\end{align}

\section{conclusion}\label{sec:conclusion}

In this work, we propose the concept of the one-way unlocalizable quantum discord and provided an operational interpretation.
We have investigated systematically some foundational properties of the one-way unlocalizable quantum discord, for example, the lower and the upper bound. By extending a bipartite system to a tripartite system, we have established several tradeoffs between the one-way unlocalizable quantum discord and other correlations.
We have also provided a polygamy relation of distributed quantum correlations in terms of the one-way unlocalizable quantum discord in tripartite quantum systems, and  have derived a lower bound for bipartite shareability of quantum correlations in a tripartite system.
If the von Neumann measurement is on a part of the system,
we have provided two expressions in terms of partial distillable entanglement and quantum disturbance for the one-way unlocalizable quantum discord.

As an important supplementary to quantum entanglement, quantum discord is proved to be also important and necessary in quantum information
processing. With recent developments for quantum correlations beyond entanglement such as quantum discord
both in experiment \cite{guonaturec} and in theory \cite{discordreview},
it will be interesting to explore the new proposed one-way unlocalizable quantum discord further experimentally and theoretically.

\begin{acknowledgments}
Y.M. Li is supported by NSFC with grant No.60873119 and the Higher School Doctoral Subject
Foundation of Ministry of Education of China with grant No.200807180005. Z.J. Xi is supported by the Superior Dissertation Foundation of Shaanxi Normal University (S2009YB03). H. Fan is supported by NSFC with grant No.10974247 and No.11175248, and `973' program (2010CB922904).
\end{acknowledgments}

\end{document}